\documentclass[aps,prl,preprint]{revtex4-1}
\usepackage{amsmath}
\usepackage{graphicx}

\begin{document}

\title{Surface roughness and thermal conductivity of semiconductor nanowires: going below the Casimir limit}

\author{J. Carrete}
\email{jesus.carrete@usc.es}

\author{L. J. Gallego}
\email{luisjavier.gallego@usc.es}

\author{L. M. Varela}
\email{luismiguel.varela@usc.es}

\affiliation{Departamento de F\'{\i}sica de la Materia Condensada, Facultad de F\'{\i}sica, Universidad de Santiago de Compostela, E-15782 Santiago de Compostela, Spain}

\author{N. Mingo}
\email{natalio.mingo@cea.fr}

\affiliation{LITEN, CEA-Grenoble, 17 rue des Martyrs, BP166, 38054 Grenoble, France}

\begin{abstract}
By explicitly considering surface roughness at the atomic level, we quantitatively show that the thermal conductivity of Si nanowires can be lower than Casimir's classical limit. However, this violation only occurs for deep surface degradation. For shallow surface roughness, the Casimir formula is shown to yield a good approximation to the phonon mean free paths and conductivity, even for nanowire diameters as thin as $2.22\;\mathrm{nm}$. Our exact treatment of roughness scattering is in stark contrast with a previously proposed perturbative approach, which is found to overpredict scattering rates by an order of magnitude. The obtained results suggest that a complete theoretical understanding of some previously published experimental results is still lacking.
\end{abstract}

\maketitle

Good thermal insulation is much harder to achieve than electrical insulation. Whereas electrical conductivity can be tuned by $14$ orders of magnitude \cite{ashcroft}, the achieved range of thermal conductivities among all known materials only spans $4$ orders of magnitude \cite{goodson}. Heat leaks easily, and a great deal of research activity is devoted to finding new ways of effectively blocking phonons \cite{cahill,pernot}. Thus, the astonishingly low thermal conductivities recently reported on Si nanowires came as a surprise \cite{nature}, since the displayed values are an order of magnitude lower than predicted by the diffuse boundary limit of Casimir's theory. Recent theoretical work has employed the Born approximation to predict a very much enhanced boundary scattering rate that would lead to a thermal conductivity well below the Casimir limit \cite{ravaioli}. However, the Born approximation is known to break down at wavelengths comparable to the size of the scatterers, so an atomic level investigation is crucial to assess whether such enhanced scattering rates are possible or not. In this Letter we present a Green's function calculation that answers the question of whether the Casimir limit to the phonon mean free path (MFP) can be overcome by large roughness. Our results show that the MFP and the thermal conductivity of a nanowire are very close to the Casimir limit if the roughness depth is less than about 20\% of the nanowire diameter. We show that the conductivity and MFP can indeed be pushed below Casimir's diffuse scattering limit, but this requires a very large surface roughening far beyond previous estimates. The interpretation of previous experimental results is revised in the light of these findings.

In 1938, H. Casimir theoretically derived the MFP for particle-like carriers in thin wires under diffuse boundary conditions \cite{casimir}. For a cylindrical wire, the direction-averaged MFP is $\lambda_{\mathrm{Casimir}}=D$, where $D$ is the wire diameter. This approach was extended by Ziman \cite{ziman} to consider the effect of partly specular surfaces, yielding $\lambda_{\mathrm{Casimir}}=\frac{1+p}{1-p}D$, where $p$ is the coefficient of specularity, ranging from $0$ for diffusive scattering to $1$ for mirror-like reflection. Work by Dingle \cite{ziman,dingle} showed that when some intrinsic bulk scattering mechanism acts together with boundary scattering, the exact solution of the Boltzmann transport equation yields nearly the same conductivity as if one used an effective MFP given by Mathiessen's approximated rule, $\lambda^{-1}\simeq\lambda^{-1}_{\text{intrinsic}}+\lambda^{-1}_{\text{Casimir}}$. For thin wires the boundary term dominates, and thus Casimir's diffuse boundary MFP determines a lower bound for the thermal conductivity.

The phonon thermal conductivity of a nanowire in the diffusive regime can be expressed \cite{mingo_calculation} as

\begin{equation}
  \kappa=\frac{k_B \omega_T}{2\pi A}\int\limits_0^\infty
  \left(\frac{\frac{\chi}{2}}{\sinh{\frac{\chi}{2}}}\right)^2
  \sum\limits_{\alpha\left(\omega\right)}
    \tau_{\alpha}\left(\omega\right)v_{\alpha}\left(\omega\right)
  d\chi,
\label{eqn:conductivity}
\end{equation}

\noindent where $k_B$ is Boltzmann's constant, $A$ the cross-sectional area of the wire, $\omega_T=k_B T/\hbar$, $\chi=\omega/\omega_T$, $\alpha$ runs over all the allowed phonon modes for each angular frequency $\omega$, and $\tau_{\alpha}$ and $v_{\alpha}$ are the relaxation time and group velocity of each mode, respectively (the product $v_{\alpha}\tau_{\alpha}$ being its MFP, $\lambda_{\alpha}$). The phononic behavior of the perfect nanowire is determined, in the harmonic approximation, by its interatomic force constant (IFC) matrix, whose spatial Fourier transform is the dynamical matrix, $\mathbf{D}$. In particular, the eigenvectors and eigenvalues of $\mathbf{D}$ are the vibrational modes of the system and their frequencies squared, respectively, and the group velocities can be obtained from the identity $2\omega v_{\alpha}=\left\langle\left. \alpha \right. \right| \frac{\partial \mathbf{D}}{\partial k} \left| \left. \alpha \right. \right\rangle$. Likewise, a particular arrangement of defects can be characterized by its perturbation matrix $\mathbf{V}$, defined as the difference between the interatomic force constant (IFC) matrices of the defective and the perfect nanowires. Its total elastic scattering cross section for an incident phonon of mode $\left| \alpha \right\rangle$ is \cite{mingo_graphene}

\begin{equation}
  \sigma_{\alpha}\left(\omega\right)=\frac{2\pi\Omega}{  \left|\left\langle\left. \alpha \right. \right| \frac{\partial \mathbf{D}}{\partial k} \left| \left. \alpha \right. \right\rangle\right|}\sum\limits_{\alpha_f} \left|\left\langle\left. \alpha_f \right.\right| \mathbf{t^+} \left|\left. \alpha \right.\right\rangle \right|^2,
  \label{eqn:crosssect}
\end{equation}

\noindent where $\Omega$ is the volume used for normalizing the wave function, $k$ the wavenumber and $\mathbf{t^+}$ is the causal $t$-matrix. This can, in turn, be  calculated from $\mathbf{V}$ and the causal Green's function of the perfect system, $\mathbf{g^+}$, using \cite{economou} the relation $\mathbf{t^+}=\left(\mathbf{1}-\mathbf{V}\mathbf{g^+}\right)^{-1}\mathbf{V}$. Cross sections and MFPs are inversely proportional through the expression $\lambda_{\alpha}^{-1}=\sigma_{\alpha}V_d^{-1}$, where $V_d^{-1}$ is the volumic density of defects. This completes the description of our approach, which can be summed up as follows: obtain the Green's function of a perfect nanowire, use it in conjunction with $\mathbf{V}$ to calculate the scattering cross section of any defect of interest for all possible frequencies and modes, convert this set of values to MFPs and substitute them in Eq. \eqref{eqn:conductivity} in order to obtain the thermal conductivity.

\begin{figure}
  \begin{center}
    \noindent\includegraphics[width=\columnwidth]{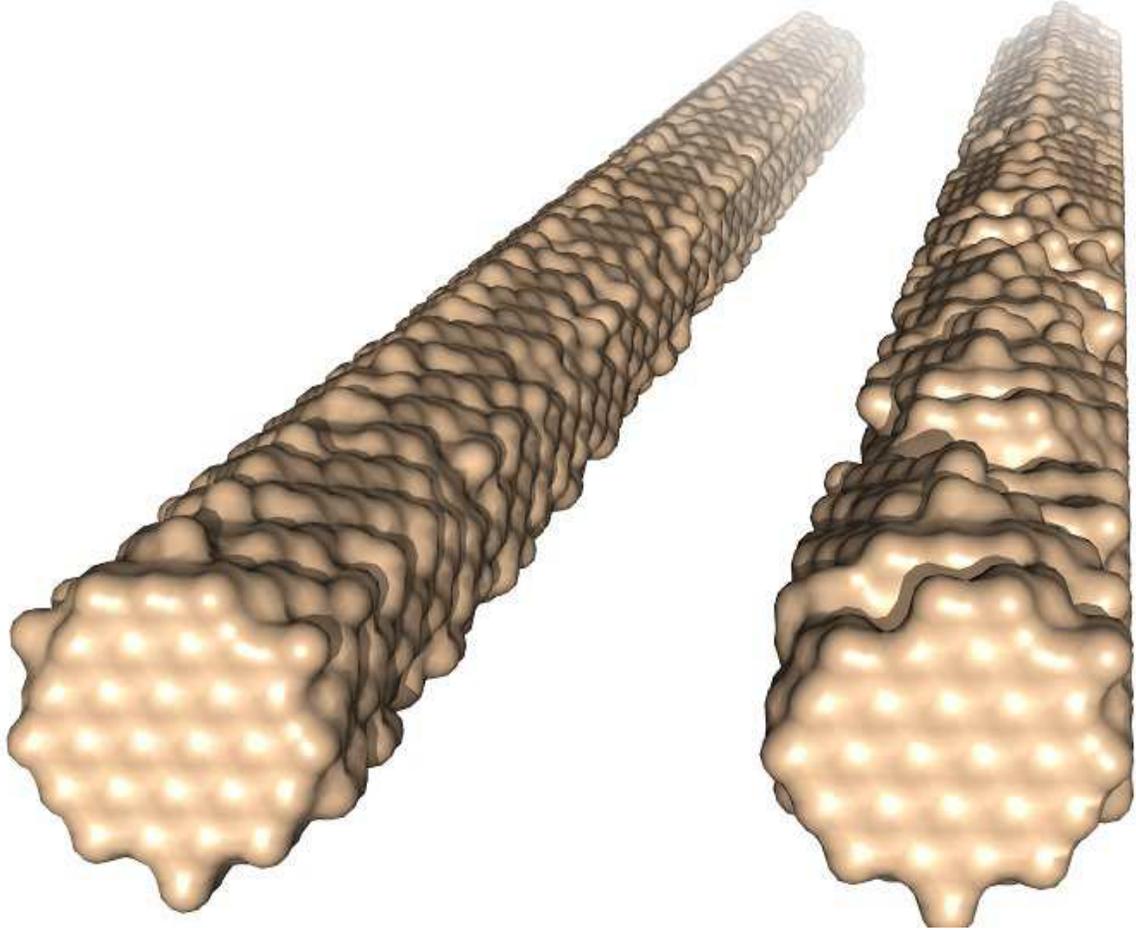}
  \end{center}
  \caption{(Color online) Left: $D=2.22\;\mathrm{nm}$ silicon nanowire taken from the $[111]$ direction of the bulk with shallow surface roughness. Right: the same nanowire with deeper cavities on its surface.
}
  \label{fig:nanowire}
\end{figure}

Experimentally, the first measurements on Si nanowires  \cite{firstmeasurements} were shown to be rather close to the diffuse boundary limit, seemingly confirming the validity of the Casimir formula  \cite{mingo_calculation}. However, since the latter is based on a semiclassical picture of phonon propagation in the structure, the question remains whether the true atomic configuration of surface imperfections might lead to different results. In order to find the answer, we have considered atomically described thin Si nanowires like the ones shown in Fig. \ref{fig:nanowire}, where surface roughness is created by removing atoms from the surface. Two scenarios have been investigated: shallow roughness, where only atoms from the outermost layer are removed at random, with 50\% probability; and deep roughness, where, on top of the existing shallow roughness, we dig deep semi-spherical cavities along the wire surface.

In this work, interactions between the atoms in the system were described by means of the Stillinger-Weber potential \cite{sw} and $\mathbf{g^+}$ was obtained using $36$ iterations of the decimation method \cite{decimation}, an iterative renormalization-group approach which, operating with the basic building blocks of the IFC matrix for the periodic nanowire, yields at its $j$-th step the relevant submatrices of the Green's function of a finite system with $2^j$ unit cells, thus quickly reaching a macroscopical, practically infinite, system. The $D_3$ symmetry of the perfect system was used in determining the normal modes at each frequency, as well as in the decimation process, by employing the incomplete projection operators \cite{groups} associated to the three irreducible representations of $D_3$ to build a symmetry-adapted basis in which the IFC, dynamical and Green's function matrices are block-diagonal. However, this symmetry is broken by the introduction of defects, and thus the sparse LU decomposition used to calculate the matrix elements of $\mathbf{t^+}$ is, by a large margin, the most time-consuming part of the calculation.

Critical in obtaining the results in Ref. \onlinecite{ravaioli} is the Born approximation $\mathbf{t^+}\simeq \mathbf{V}$. This first-order perturbative approximation lowers the computational costs of the calculation dramatically and was also tried in this work in order to compare its results with the exact ones.

\begin{figure}
  \begin{center}
    \noindent\includegraphics[width=\columnwidth]{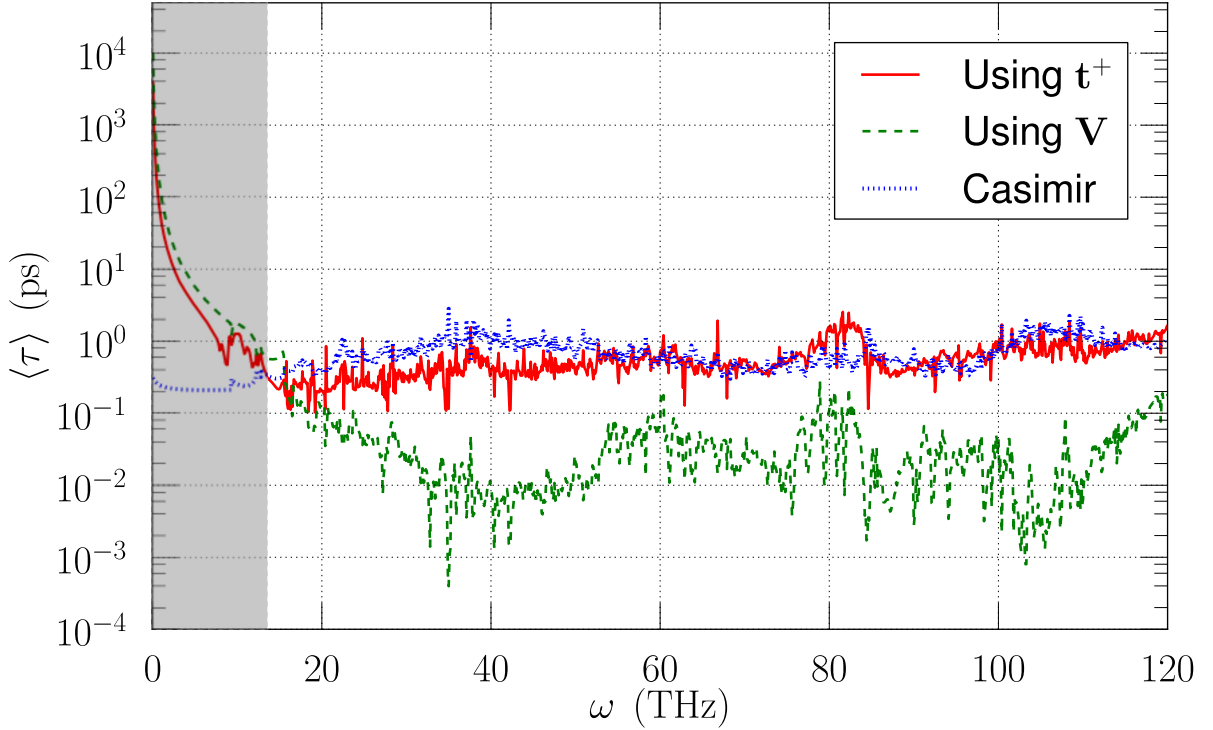}
  \end{center}
  \caption{(Color online) Relaxation times, averaged over all the available modes for each frequency, in a $D=2.22\;\mathrm{nm}$ wire with half of its outermost atoms removed, calculated using the exact $\mathbf{t^+}$ matrix and the Born approximation. The Casimir limit is also plotted for comparison. The shadowed area corresponds to the part of the spectrum for which the resistance of the contacts would be the dominating factor, as explained in the text.}
  \label{fig:tau}
\end{figure}

Using a nanowire with a diameter $D=2.22\;\mathrm{nm}$ (four atomic layers in the radial direction, 182 atoms in the unit cell) as the starting perfect system, we removed half of its outermost atoms at random to emulate surface disorder. The associated scattering cross sections were then computed using Eq. \eqref{eqn:crosssect}, and the MFPs and relaxation times were derived from them. The average relaxation times for each frequency are shown in Fig. \ref{fig:tau}. Both the exact results and those obtained using the Born approximation diverge at low frequencies, indicating quasi-ballistic transmission. At higher frequencies, the exact relaxation times suggest that the Casimir approximation is adequate. There is, however, a striking difference of $1-2$ orders of magnitude between the red and green curves, with the Born approximation resulting in an unrealistically high estimate of the scattering cross section. This is a general feature of the approximation in this context and becomes even worse when larger defects are introduced. Further evidence of this failure is the fact that, if the total reflectance of a single defective segment is calculated using \cite{1d}

\begin{equation}
  R\left(\omega\right)=\sum\limits_{\alpha\left(\omega\right),\alpha_f\left(\omega\right)}
  \frac{\left| \left\langle \alpha_f^* \left| \mathbf{t^{+}} \right| \alpha \right\rangle\right|^2}
{\left|
    \left\langle\left. \alpha_f \right.\right| \frac{\partial\mathbf{D}}{\partial k} \left|\left. \alpha_f \right.\right\rangle
\right|\left|
  \left\langle\left. \alpha \right.\right| \frac{\partial\mathbf{D}}{\partial k} \left|\left. \alpha \right.\right\rangle
\right|},
\label{eqn:R}
\end{equation}

\noindent and again introducing the perturbative approximation $\mathbf{t^+}\simeq\mathbf{V}$, for most frequencies this reflectance is much larger than the transmission of the perfect system. This is a clearly absurd result which, if substituted into the Landauer formula \cite{datta} and integrated, would yield a negative value for the thermal conductivity.

Surface roughness with a characteristic size larger than the size of the simulation cell ($1.9\;\mathrm{nm}$) is not included in the simulation and thus the scattering of longer-wavelength/low-frequency phonons is Rayleigh-like, nearly ballistic. The contribution of this part of the spectrum to the thermal conductivity is, however, negligible if we take into account the frequency dependence of the transmission probability between the contacts used for measuring the thermal conductance and the nanowire itself, which is proportional to $\omega^2$, as shown in Ref. \onlinecite{contacts}. This probability was included in our thermal conductivity calculation using a composition rule for the transmission of the wire plus contact system \cite{chandan}. The region where contact scattering dominates the composition, causing the results from $t$-matrix calculations alone to be insufficient, is shown shadowed in Fig. \ref{fig:tau}.

\begin{figure}
  \begin{center}
    \noindent\includegraphics[width=\columnwidth]{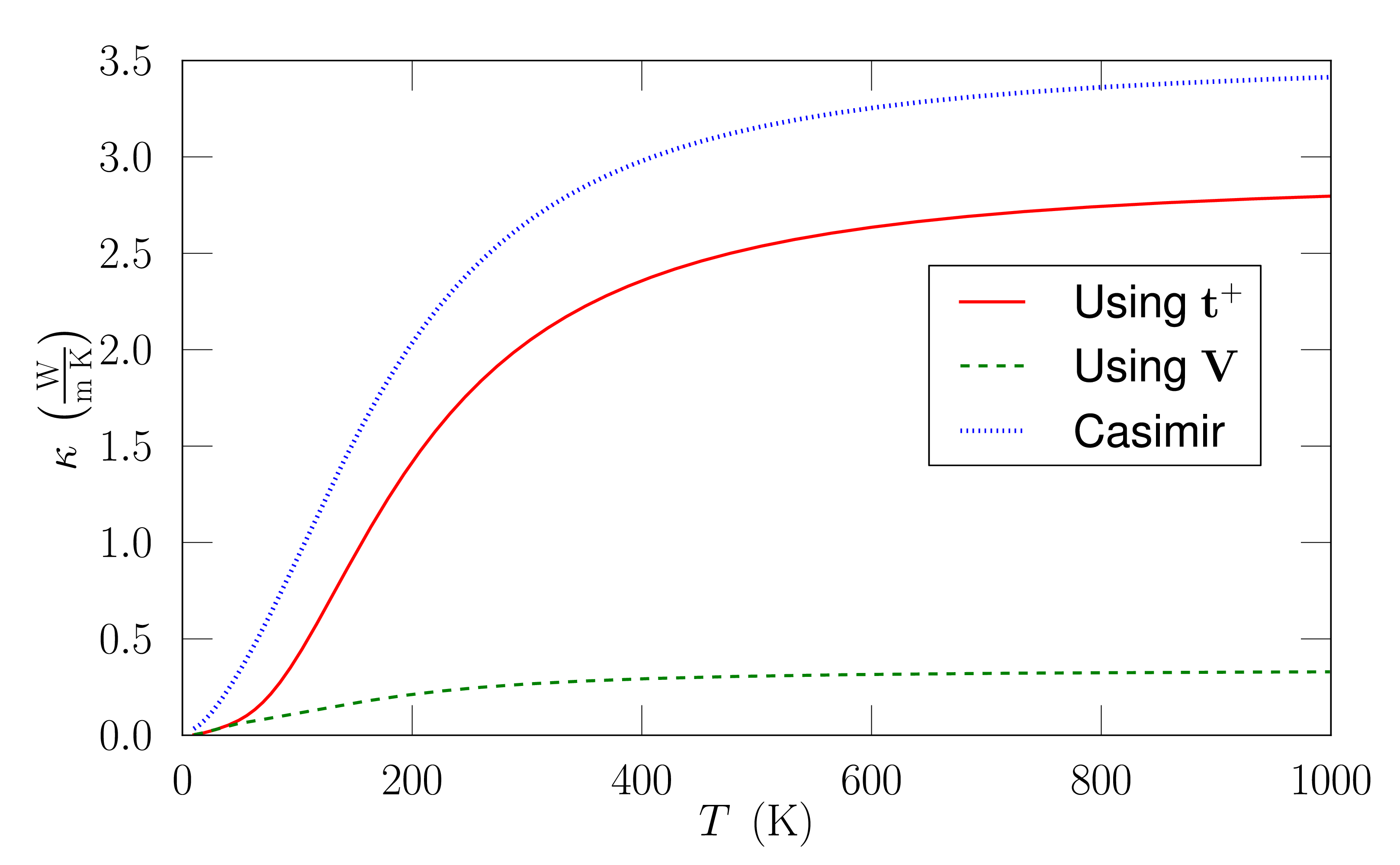}
  \end{center}
  \caption{(Color online) Thermal conductivity of a $D=2.22\;\mathrm{nm}$ nanowire with half of its outermost atoms removed, as a function of temperature, calculated using Eq. \eqref{eqn:conductivity} either with the exact $t$-matrix, the Born approximation or the Casimir MFP.}
  \label{fig:conductivity}
\end{figure}

Figure \ref{fig:conductivity} compares the thermal conductivities calculated from Eq. \eqref{eqn:conductivity} using MFPs predicted by Eq. \eqref{eqn:crosssect} ---in red---, the Born approximation ---dashed green---  or the Casimir formula \footnote{For a non-cylindrical nanowire with translational symmetry, $\lambda_{\mathrm{Casimir}}=FD$, where $F$ is a geometric form factor (Z. Wang and N. Mingo, unpublished). For hexagonal nanowires, $F=0.907$. We used a weighted value of the diameter to take the missing atoms into account.} ---dotted blue---  for the system discussed in the foregoing paragraph. The exact MFPs afford a thermal conductivity in reasonable agreement with the Casimir approximation, whereas the perturbative result is qualitatively and quantitatively different, reaching saturation at much lower temperature and conductivity. The values shown here agree in order of magnitude with those from MD simulations of rough Si nanowires reported in Ref. \onlinecite{donadio}. There could be some overestimation of the conductivities due to the fact that the Stillinger-Weber potential yields phonon frequencies above their true values, but this would affect the three curves in Fig. \ref{fig:conductivity} and thus not undermine our conclusions. Since very thin nanowires are those most removed from the context of the derivation of Casimir's formula (which is known to be valid for thick ones), our results suggest that this formula should be a good approximation also for all larger thicknesses. The role of the contact-nanowire transmission (less important for thick wires) in suppressing the divergence at low frequencies would be played in that case by the vanishing, bulk-like number of available states for each frequency \cite{mingo_calculation}.

\begin{figure}
  \begin{center}
    \noindent\includegraphics[width=\columnwidth]{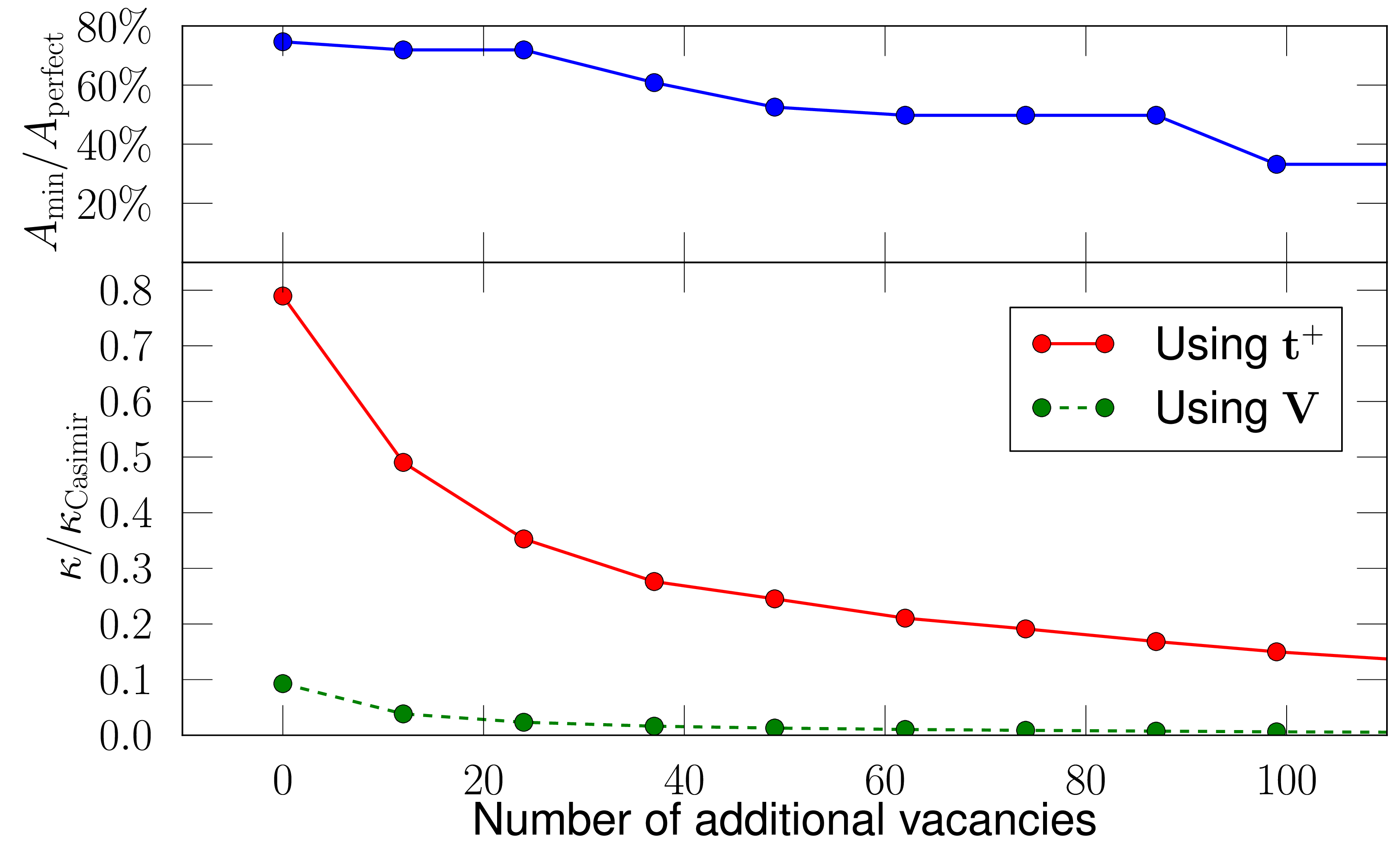}
  \end{center}
  \caption{(Color online) Bottom: Thermal conductivity at $T=1000\;\mathrm{K}$, relative to the Casimir value, of a $D=2.22\;\mathrm{nm}$ nanowire with half of its outermost atoms and an additional cluster of atoms removed, as a function of temperature, calculated using Eq. \eqref{eqn:conductivity} with the exact $t$-matrix (solid line) or the Born approximation (dashed line). Top: minimum cross-sectional area of each nanowire, relative to the perfect system.}
  \label{fig:cluster}
\end{figure}

It is possible to go well below the Casimir value by blocking the phonons using defects which penetrate deeper into the nanowire. Figure \ref{fig:cluster} shows the effect on the conductivity (computed at $T=1000\;\mathrm{K}$) of removing an almost semi-spherical cluster of up to $100$ atoms from each $1.9\;\mathrm{nm}$-long segment of the wire. The conductivity is normalized to the value obtained by the Casimir formula. As before, the Born approximation overestimates the thermal resistance by an order of magnitude. We can conclude from the $t$-matrix results that the reduction in conductivity resulting from further removal of atoms is rather slow once about a quarter of the cross-sectional area of the wire has been removed. Thus, it is difficult to achieve a reduction in conductivity of more than one order of magnitude with respect to the Casimir limit without compromising other desirable properties or even the structural stability of the system.

The progressively smaller efficiency of further removal of atoms is to be expected if we compare the situation, qualitatively, to the scattering of phonons by spherical nanoparticles studied by Kim and Majumdar \cite{mie}, who suggest an interpolation formula of the form $\sigma^{-1}=\sigma_{\mathrm{Rayleigh}}^{-1}+\sigma_{\mathrm{Near\; gometrical}}^{-1}$, resulting in a scattering efficiency that increases quickly with the radius of the sphere when it is small, but simply oscillates around the geometrical limit for larger radii.

The possibility of reducing wire conductivity below the Casimir limit had been investigated in the past using non-atomistic approaches. The Montercarlo work by Moore \textit{et al.} \cite{sawtooth} introduced deep boundary roughness in the form of sawtooth boundaries. They found that phonon backscattering can cause the coefficients of specularity to become negative, indicating MFPs below Casimir's limit. However, neither their predictions nor the ones reported in this Letter match the extremely low experimental conductivities found in Ref. \onlinecite{nature}. This is especially surprising since the depth of the disordered layers in the experimental nanowires is smaller (relative to their diameter) than in the simulations. All of this and the difference in order of magnitude between the exact and approximate curves in Fig. \ref{fig:cluster} suggest that the interpretation of experimental data in Ref. \onlinecite{ravaioli} may be just an artifact caused by the Born approximation.

In summary, starting with an atomistic model and performing exact $t$-matrix calculations, we have found that the Casimir limit can be overcome. However, by quantifying the magnitude of this violation in terms of surface roughness, we have shown that in most feasible situations the Casimir formula still constitutes a good approximation to the effect of boundary scattering on phonon transport in thin semiconductor nanowires with surface disorder and that, as a result, achieving a reduction in conductivity of more than one order of magnitude below Casimir's limit might involve such a strong surface distortion so as to compromise the structural stability of the nanowire. In particular, our results indicate that surface roughness cannot be the only cause for the extremely low thermal conductivity that has been experimentally reported for Si nanowires \cite{nature}.

\begin{acknowledgments}
  This work was supported by France's Agence Nationale de la Recherche, Fondation Nanosciences, the Spanish MICINN/FEDER (FIS2008-04894/FIS) and the Xunta de Galicia (INCITE09E2R206033ES). J. C. thanks the Spanish Ministry of Education for a FPU grant.
\end{acknowledgments}

\end{document}